\documentclass[preprintnumbers,floatfix,9pt,prd,superscriptaddress,nofootinbib,onecolumn]{revtex4}
\usepackage{graphicx}
\usepackage{epsfig}
\usepackage{bm}
\usepackage{amsfonts}

\begin{document}

\title{Black hole solutions in $F(R)$ gravity with conformal anomaly}
\author{\textbf{S. H. Hendi}}
\email{hendi@mail.yu.ac.ir}
\affiliation{Physics Department, College of Sciences, Yasouj University, Yasouj 75914,
Iran.}
\author{\textbf{D. Momeni}}
\email{d.momeni@yahoo.com} \affiliation{Department of Physics,
Faculty of sciences, Tarbiat Moa'llem university, Tehran, Iran}

\begin{abstract}
\vspace*{0.3cm} \centerline{\bf Abstract} \vspace*{0.3cm}

In this paper, we consider $F(R)=R+f(R)$ theory instead of
Einstein gravity with conformal anomaly and look for its
analytical solutions. Depending on the free parameters, one may
obtain both uncharged and charged solutions for some classes of
$F(R)$ models. Calculation of Kretschmann scalar shows that there
is a singularity located at $r=0$, which the geometry of uncharged
(charged) solution is corresponding to the Schwarzschild (Reissner-Nordstr%
\"{o}m) singularity. Further, we discuss the viability of our models in
details. We show that these models can be stable depending on their
parameters and in different epoches of the universe.
\end{abstract}

\maketitle

\section{Introduction}

In $1970$, Buchdahl considered the first modification of the
Einstein Lagrangian density involving an arbitrary function of
Ricci scalar, $F(R)$ \cite{Buchdahl1970}. After that and motivated
by inflationary scenarios, Starobinsky proposed an action which is
to weigh up the effects of $R^{2}$ corrections to Einstein gravity
\cite{Starobinsky1980}. In addition, the recent advent of new
observational precision tests, such as rotation curves of spiral
galaxies \cite{RotationCurves} and solar system tests
\cite{SolarSystem} have changed the modern view of cosmos based on
General Relativity. Amongst the modification of Einstein gravity,
the so-called $F(R)$ gravity is completely special (see
\cite{F(R)} and references therein and for a review see
\cite{F(R)Rev}). Interpretation of the hierarchy and singularity
problems \cite{HProb,SProb1,SProb2}, early-time inflation
\cite{ET}, gravitational waves detection \cite{GW} and also the
four cosmological phases \cite{FCP}, have been investigated in
$F(R)$ models.

On the other side, the conformal anomaly (CA) \cite{CA,anomaly}
plays a crucial role in string theory \cite{Polyakov} and
gravitational field theory \cite{Parker}. Also, it has been shown
that there is a deep relation between Hawking radiation and trace
anomaly \cite{HRTA}. The relation of CA with some cosmological
problems such as inflation and cosmological constant problem have
been considered in Refs. \cite{ICCP,anomaly}. Applications of CA
in black hole and particle physics have been studied in
\cite{CA1,CA2,CaiCA} and \cite{ParCA}, respectively (for a
suitable treatment of CA in the black hole solutions see
\cite{CA2}). In addition, we should note that the first
Schwarzschild-dS solutions in $F(R)$ gravity has been discussed in
\cite{SdS}.

In the semiclassical framework of general relativity, one may use
quantum field theory (QFT) to describe the matter while the
gravitational field is considered as a classical field. When
classical gravity is regarded as a background for a conformally
coupled QFT, the trace of the classical expectation value of
matter field energy momentum tensor does not vanish. In other
word, considering the quantum expectation value of the energy
momentum tensor during the process of renormalization, leads to
additional terms in the Einstein-Hilbert action.

It has been shown that, when $F''(R) \neq 0$, metric $F(R)$
gravity is completely equivalent to a special case of
scalar-tensor theory \cite{STF(R)}. In order to find a
scalar-tensor representation of $F(R)$ theories, one can use
conformal transformations \cite{STF(R)}. Considering suitable
conformal transformations, one may find that, $F(R) = R+f(R)$
behaves geometrically as if we have $F(R) = R+$ (non-minimal
scalar field). Remarkably, $f(R)$ plays the role of "scalar field"
so that the geometry $F(R) = R+f(R)$ becomes equivalent to the
Einstein-dilaton geometry in a spherically symmetry spacetime. It
is obvious that the CA may come from the quantum effects of matter
field. Regarding the scalar-tensor representation of $F(R)=R+f(R)$
theories, one may consider $f(R)$ (or equivalently "scalar field")
as a matter field.

It is notable that there are in fact two distinct types of trace
anomalies with different geometric and physical antecedents. The
first anomaly called type A anomaly which is related to the Euler
characteristic of $2n$-dimensional spacetime and the second named
B anomaly with a connection to the Weyl tensor.

In this paper, we consider a class of black hole in $F(R)$ gravity
with trace anomaly and look for analytical solutions. In order to
study the basics solutions of general $F(R)$ gravity theories with
constant curvature solutions $R=R_{0}$ with CA, let us start from
the action
\begin{equation}
S=S_{g}+S_{m},  \label{S}
\end{equation}%
in which $S_{g}$ is the $4$-dimensional action

\begin{equation}
S_{g}=\frac{1}{16\pi G}\int d^{4}x\sqrt{-g}\left[ R+f(R)\right] ,  \label{Sg}
\end{equation}%
where $8\pi G=M_{p}^{-2}$, $M_{p}$ corresponds with the Planck
mass, $g$ is determinant of the metric $g_{\mu \nu }(\mu ,\nu
=0,1,2,3)$, $R$ is the scalar curvature and $R+f(R)$ is the
function defining the theory under consideration. As the simplest
example, the Einstein gravity with cosmological constant $\Lambda
$ is given by $f(R)=-2\Lambda $. We know that one loop quantum
corrections lead to a trace anomaly of the energy-momentum (EM)
tensor of conformal field theories. In general, as we can show
that, the trace anomaly has the form \cite{anomaly}
\begin{equation}
<T_{\mu \nu }>=g_{\mu \nu }\left[ \beta C_{\alpha \beta \gamma \delta
}C^{\alpha \beta \gamma \delta }-\alpha (R^{2}-4R_{\alpha \beta }R^{\alpha
\beta }+R_{\alpha \beta \gamma \delta }R^{\alpha \beta \gamma \delta })%
\right] ,  \label{T1}
\end{equation}%
where $\alpha ,\beta $ are two positive constants depending on the degrees
of freedom of quantum fields and their values are not important for our
discussions. The first term is a polynomial of Weyl tensor (B anomaly),
while the second (Gauss-Bonnet) term is Euler characteristic of the $4$%
-dimensional spacetime (A anomaly). From the above considerations, the
equations of motion (EOM) in the metric formalism are just
\begin{eqnarray}
&&R_{\mu \nu }\left( 1+f_{R}\right) -\frac{1}{2}\left[ R+f(R)\right] g_{\mu
\nu }+(\nabla _{\mu }\nabla _{\nu }-g_{\mu \nu }\nabla _{\mu }\nabla ^{\mu
})f_{R}  \nonumber \\
&&-8\pi Gg_{\mu \nu }\left[ \beta C_{\alpha \beta \gamma \delta }C^{\alpha
\beta \gamma \delta }-\alpha (R^{2}-4R_{\alpha \beta }R^{\alpha \beta
}+R_{\alpha \beta \gamma \delta }R^{\alpha \beta \gamma \delta })\right] =0,
\label{FE1}
\end{eqnarray}%
where $f_{R}=\frac{df(R)}{dR}$. The problem of finding the general static
spherically symmetric (SSS) solution in arbitrary $f(R)$ theories without
imposing the constant curvature condition is in principle too complicated.
The required condition to get vacuum constant curvature solutions $R=R_{0}$%
(from now $R_{0}$ will denote a constant curvature value) in vacuum implies
\begin{equation}
R_{0}\left( 1+f_{R_{0}}\right) =2\left[ R_{0}+f(R_{0})\right]  \label{VCC}
\end{equation}%
For this kind of solutions, an effective cosmological constant may
be defined as $\Lambda _{D}^{eff}=\frac{R_{0}}{4}$. This type of
black-hole discussed in \cite{bh}. It has been proved that the
only static and spherically symmetric vacuum solution (SSVS) with
constant curvature of any $F(R)$ gravity is just the Hawking- Page
black hole in AdS space \cite{HP}. But for the theory with trace
anomaly with $T=T_{\mu }^{\mu }$ this condition is
\begin{equation}
R\left( 1+f_{R}\right) -2\left[ R+f(R)\right] -8\pi GT=0  \label{TraceFE1}
\end{equation}%
One can consider (\ref{TraceFE1}) as a differential equation (DE) for the $%
f(R)$ function and with a $T=T(r)\equiv T(R)$. The general solution for DE (%
\ref{TraceFE1}) is given by
\begin{equation}
f(R)=-R+c_{1}R^{2}+8\pi GR^{2}\int^{R}\frac{T(R^{\prime })}{R^{\prime 3}}%
dR^{\prime }  \label{Generalf(R)}
\end{equation}

\section{Assumption of EM tensor with topological metric}

We consider the external metric for the gravitational field produced by a
non rotating object in $f(R)$-Weyl (CA) gravity. The general $%
4$-dimensional topological metric can be written as \cite{SSS}
\begin{equation}
ds^{2}=\lambda (r)dt^{2}-\mu ^{-1}(r)dr^{2}-r^{2}\left( d\theta
_{1}^{2}+\Omega ^{2}d\theta _{1}^{2}\right) .  \label{Metric}
\end{equation}%
where $\Omega $ denotes $\sin \theta _{1}$, $1$ and $\sinh \theta _{1}$ for $%
k=1$ (spherical horizon), $k=0$ (flat horizon), and $k=-1$ (hyperbolic
horizon), respectively. We follow the methodology of Cai et.al \cite{CaiCA}.
We making an assumption that to have a non trivial black hole solution with
metric (\ref{Metric}) at a black hole horizon, say $r_{H}$, the EM tensor
must satisfy the auxiliary condition, $%
<T_{t}^{t}(r)>_{r=r_{H}}=<T_{r}^{r}(r)>_{r=r_{H}}$, which holds not on
horizon even in the whole spacetime. We take the EM tensor for SSS
configurations as
\begin{equation}
T_{\mu }^{\nu }=diag(\rho (r),-p(r),-p_{\bot }(r),-p_{\bot }(r)),  \label{T2}
\end{equation}%
where its trace is
\begin{equation}
T=\rho (r)-p(r)-2p_{\bot }.  \label{TraceT}
\end{equation}%
Applying the auxiliary condition to metric (\ref{Metric}), one can easily
show that $\lambda (r)=\mu (r)$. Our main goal is searching for an exact
solution for (\ref{FE1}) with definite $F(R)$ action given by (\ref%
{Generalf(R)}) and determining the metric functions of metric (\ref{Metric}%
). The trace $T$ from Eq. (\ref{TraceT}) is not constant since the
components of the EM tensor (\ref{T2}) are functions of the radial
coordinate $r$. Thus in general, the $f(R)$ functions are very different
depending on the form of the $T$ from (\ref{TraceT}). Locally, let us to
take $T$ as a constant which we can find that Eq. (\ref{Generalf(R)})
reduces to
\begin{equation}
f(R)=-R+c_{1}R^{2}-4\pi GT.  \label{f(R)1}
\end{equation}%
This action is a second order corrected action of the Einstein-Hilbert
action with a cosmological constant $\Lambda =2\pi GT$. Thus it's solution
is a second order corrected (anti) de Sitter spacetime for a typical
(negative) positive value of $T$. Since the Lagrangian of Einstein gravity ($%
R$-term) remove in this model (\ref{f(R)1}), the corresponding solutions are
not important for us. In next section we will convert the Eq. (\ref{FE1}) to
a more convent form.

\section{Near horizon and approximate solutions for constant curvature}

In this section we focuss on the special case $f(R)=-2\Lambda $, with
constant curvature i.e. $R=R_{0}$. Firstly it is adequate to rewrite the EM (%
\ref{T1}) as the following form
\begin{equation}
<T_{\mu \nu }>=g_{\mu \nu }\left[ \left( \frac{\beta }{3}-\alpha \right)
R^{2}-2\left( \beta -2\alpha \right) R_{\alpha \beta }R^{\alpha \beta
}+(\beta -\alpha )R_{\alpha \beta \gamma \delta }R^{\alpha \beta \gamma
\delta })\right] ,  \label{T11}
\end{equation}%
where we used the following identity \cite{invar}

\begin{equation}
C_{\alpha \beta \gamma \delta }C^{\alpha \beta \gamma \delta }=\frac{1}{3}%
R^{2}-2R_{\alpha \beta }R^{\alpha \beta }+R_{\alpha \beta \gamma \delta
}R^{\alpha \beta \gamma \delta }.  \label{CC}
\end{equation}%
As we pointed former, $\alpha ,\beta $ are two constants depending on the
degrees of freedom of quantum fields and their values are not important for
our discussion. We choice $\alpha =\beta $ and also define a new parameter $%
\alpha ^{\prime }=2\alpha $. Thus, we write the next EOM
\begin{eqnarray}
&&R_{\mu \nu }(1+f_{R})-\left( \frac{1}{2}\left[ R+f(R)\right] -\frac{8\pi
G\alpha ^{\prime }}{3}R^{2}\right) g_{\mu \nu }  \nonumber \\
&&+(\nabla _{\mu }\nabla _{\nu }-g_{\mu \nu }\nabla _{\beta }\nabla ^{\beta
})f_{R}=8\pi G\alpha ^{\prime }g_{\mu \nu }(R_{\alpha \beta }R^{\alpha \beta
}).  \label{FE2}
\end{eqnarray}%
Constant curvature solutions for (\ref{FE2}) with $f(R)=-2\Lambda $, can be
written as
\begin{equation}
R_{\mu \nu }=\alpha ^{\prime \prime }g_{\mu \nu }(\gamma +R_{\alpha \beta
}R^{\alpha \beta }),  \label{Ricci}
\end{equation}%
where $\gamma =\frac{\frac{1}{2}R_{0}-\frac{1}{3}\alpha ^{\prime \prime
}R_{0}^{2}-\Lambda }{\alpha ^{\prime \prime }}$ and $\alpha ^{\prime \prime
}=8\pi G\alpha ^{\prime }$. Applying the metric (\ref{Metric}) with $\lambda
=\mu $ to Eq. (\ref{Ricci}), one leads to the next nonlinear differential
equations (NDEs)

\begin{eqnarray}
\lambda ^{\prime \prime }(r) &=&\frac{1}{r^{2}}\left[ -2r\lambda ^{\prime
}(r)\pm \sqrt{2r^{4}\left[ -\gamma +\frac{1}{8\alpha ^{\prime \prime 2}}%
\right] -4\left[ r\lambda ^{\prime }(r)+\lambda (r)-k\right] ^{2}}\right] -%
\frac{1}{2\alpha ^{\prime \prime }}  \label{lambdaPP1} \\
\lambda ^{\prime \prime }(r) &=&\frac{1}{r^{2}}\left[ -2r\lambda ^{\prime
}(r)\pm \sqrt{2r^{4}\left[ -\gamma +\frac{-r\lambda ^{\prime }(r)+k-\lambda
(r)}{r^{2}\alpha ^{\prime \prime }}\right] -4\left[ r\lambda ^{\prime
}(r)+\lambda (r)-k\right] ^{2}}\right]   \label{lambdaPP11}
\end{eqnarray}%
where prime denotes the derivative with respect to the radial coordinate $r$%
. We must solve Eqs. (\ref{lambdaPP1}) and (\ref{lambdaPP11})
simultaneously. This NDEs have no (simple) exact solution but we can solve
it approximately or numerically with a suitable boundary conditions.
According to the Hawking-Bekenstein formula for temperature of the black
hole, if the metric (\ref{Metric}) posses a black hole solution with a
horizon located at $r=h$, we can deduce $\lambda (h)=0$ and

\begin{equation}
T=\frac{\lambda ^{\prime }(h)}{4\pi }.  \label{T3}
\end{equation}%
Examining (\ref{lambdaPP1}) and (\ref{lambdaPP11}) at horizon we can write

\begin{eqnarray}
\lambda ^{\prime \prime }(h) &=&\frac{1}{h^{2}}\left[ -8\pi hT\pm \sqrt{%
2h^{4}\left[ -\gamma +\frac{1}{8\alpha ^{\prime \prime 2}}\right] -4(4\pi
hT-k)^{2}}\right] -\frac{1}{2\alpha ^{\prime \prime }}  \label{Lam1} \\
\lambda ^{\prime \prime }(h) &=&\frac{1}{h^{2}}\left[ -8\pi hT\pm \sqrt{%
2h^{4}\left[ -\gamma +\frac{-4\pi hT+k}{\alpha ^{\prime \prime }h^{2}}\right]
-4(4\pi hT-k)^{2}}\right]   \label{Lam2}
\end{eqnarray}

Here, we desire to write near horizon solution of $\lambda (r)$. To do this,
we equate Eq. (\ref{lambdaPP1}) with Eq. (\ref{lambdaPP11}) and obtain

\begin{equation}
\alpha ^{\prime \prime }=\alpha _{c}=-\frac{3h^{2}\left[ 8\pi hT-2\Lambda
h^{2}+R_{0}h^{2}-2k\right] }{2\left[ 12\left( 4\pi hT-k\right)
^{2}-R_{0}^{2}h^{4}\right] }.  \label{alphapp}
\end{equation}%
In other word, Eq. (\ref{lambdaPP1}) is equal to Eq. (\ref{lambdaPP11}) for
specific value of $\alpha ^{\prime \prime }$, (\ref{alphapp}). Thus, the
near horizon solution of $\lambda (r)$ of Eqs. (\ref{lambdaPP1}) and (\ref%
{lambdaPP11}) with Eq. (\ref{alphapp}) is obtained as
\begin{equation}
\lambda (r)\simeq 4\pi T(r-h)+\frac{(r-h)^{2}}{2h^{2}}\left[ -8\pi hT\pm
\sqrt{2h^{4}\left[ -\gamma +\frac{-4\pi hT+k}{\alpha _{c}h^{2}}\right]
-4(4\pi hT-k)^{2}}\right] .  \label{lambda1}
\end{equation}

On the other side, for small values of $\alpha ^{\prime \prime }$, we can
expand Eqs. (\ref{lambdaPP1}) and (\ref{lambdaPP11}) and solve the resulting
DEs with the following solution
\begin{equation}
\lambda (r)=k-\frac{2M}{r}+\frac{r^{2}}{6}\left( 2\Lambda -R_{0}\right) .
\label{lambda2}
\end{equation}%
Identifying this metric function with a Schwarzschild-(a)ds metric gives
\begin{equation}
R_{0}=4\Lambda ,  \label{Constant}
\end{equation}%
and thus, for small values of $\alpha ^{\prime \prime }$, the metric (\ref%
{Metric}) reads as
\begin{equation}
ds^{2}=\left[ k-\frac{2M}{r}+\frac{r^{2}}{6}\left( 2\Lambda -R_{0}\right) %
\right] dt^{2}-\frac{dr^{2}}{\left[ k-\frac{2M}{r}+\frac{r^{2}}{6}\left(
2\Lambda -R_{0}\right) \right] }-r^{2}d\Omega ^{2}  \label{Metric2}
\end{equation}

\section{Exact solutions}

\subsection{$f(R)=-2\Lambda $ model:}

\subsubsection{Case I: $\protect\alpha \neq 0$, $\protect\beta =0$}

At the first step and following the procedure in \cite{CaiCA}, we consider $%
f(R)=-2\Lambda $ with type A anomaly (Gauss-Bonnet term). It is easy to show
that for $\lambda (r)=\mu (r)$, the field equation (\ref{FE1}) reduces to
\begin{equation}
\frac{d}{dr}\left[ \Pi +\Pi _{ii}\right] =0,  \label{diffEq}
\end{equation}%
where%
\begin{eqnarray}
\Pi &=&2\alpha ^{\prime \prime }\lambda ^{\prime }(r)\left[ \lambda (r)-k%
\right] +\frac{\Lambda r^{3}}{3},  \label{PI} \\
\Pi _{ii} &=&\left\{
\begin{array}{cc}
r\left[ \lambda (r)-k\right] & i=t,r \\
\frac{r^{2}}{2}\lambda ^{\prime }(r) & i=\theta ,\phi%
\end{array}%
\right. ,  \label{PIii}
\end{eqnarray}%
Using Eq. (\ref{diffEq}) with Eqs. (\ref{PI}) and (\ref{PIii}), the metric
function can be obtained as
\begin{equation}
\lambda (r)=k-Ar^{2},  \label{L1}
\end{equation}%
where
\begin{equation}
A=\frac{3\pm \sqrt{9-48\Lambda \alpha ^{\prime \prime }}}{24\alpha ^{\prime
\prime }}.  \label{A}
\end{equation}%
We should note that for $\alpha ^{\prime \prime }=3/(16\Lambda )$, Eq. (\ref%
{A}) reduces to $2\Lambda /3$ and the presented solution is close to
asymptotic (a)dS solution.

\subsubsection{Case III: arbitrary $\protect\alpha =\protect\beta $}

In what follows, we obtain an exact solution of the field equation
(\ref{lambdaPP1}) with A and B anomalies, simultaneously with
$f(R)=-2\Lambda $, which is Einstein-$\Lambda $ gravity with CA.
Straightforward calculations show that, in this model, one can
obtain topological Schwarzschild solution as \cite{SdS}
\begin{equation}
\lambda (r)=\mu (r)=k-Ar^{2}-\frac{2M}{r},  \label{Sch}
\end{equation}
where $A$ is given in Eq. (\ref{A}) and $\beta ^{\prime \prime }=16\pi
G\beta =\alpha ^{\prime \prime }.$ As we mentioned before, one can obtain
approximate Schwarzschild-(a)dS solution, provided the parameters of the
solution are chosen suitably.

\subsection{$f(R)=c_{1}R^{2}-2\Lambda $ model:}

Considering a linear function of $T(R^{\prime })=cR^{\prime }+b$, we find
that Eq. (\ref{Generalf(R)}) leads to the Einstein-Hilbert action with a $%
R^{2}$ correction with a cosmological constant $\Lambda =2b\pi G$ and $c=%
\frac{-1}{8\pi G}$. Straightforward calculations show that Eq. (\ref{Sch})
is a solution of field equation (\ref{FE1}) for $\alpha ^{\prime \prime
}=\beta ^{\prime \prime }$ and obtained $A$ (\ref{A}).

We note that asymptotic flat Schwarzschild solution may be found by setting $%
k=1$ and $b=0$, which exactly yields the vanishing cosmological constant. In
other word, $R^{2}$ correction does not effect on the asymptotic behavior of
the solutions.

In addition to the Schwarzschild solution, one can show that
topological charged solution is another exact solution of this
model. In other word, we should note that
\begin{equation}
\lambda (r)=\mu (r)=k-Ar^{2}+\frac{q^{2}}{r^{2}},  \label{ChargeSolution}
\end{equation}
is a solution of FE (\ref{FE1}), provided the parameters of the
solution are chosen as follow
\begin{eqnarray}
\beta ^{\prime \prime } &=&\frac{5\alpha ^{\prime \prime }}{6},
\label{beta1} \\
c_{1} &=&\frac{-1}{24A},  \label{c1}
\end{eqnarray}
where $A$ is the same as Eq. (\ref{A}).

It is notable that the presented uncharged black holes is the same
as solutions which obtained in Ref. \cite{SZ} for $\alpha=0$.
Although for $g_{tt}=g^{-1}_{rr}$ in Ref. \cite{SZ}, the metric
function has a charged term $\frac{C_{1}}{r^{2}}$ in analogy with
our charged solution, but these solutions are completely
different. They have different geometry and Ricci scalar $R$ with
various asymptotic behavior.

Here, we want to discuss about the stability of such solution as
an example of $F(R)$ gravity solutions with CA. The stability of
the solutions in F(R) gravity has been discussed in literatures
\cite{Stab1}. In fact, the stability of the de Sitter solution may
be obtained by imposing the one-loop effective action to be real.
In order to study of stability, some various techniques have been
employed which all these methods are in agreement with the
following conditions which state the existence and the stability
of the de Sitter solution:
\begin{eqnarray}
I &:&2F(R_{0})=R_{0}F^{\prime }(R_{0}),  \nonumber \\
II &:&\frac{F(R_{0})}{F^{\prime }(R_{0})}>0,  \nonumber \\
III &:&\frac{F^{\prime }(R_{0})}{R_{0}F^{\prime \prime
}(R_{0})}>1. \label{StabCond}
\end{eqnarray}
Equations $I$ and $II$, state the existence of a solution with
positive constant curvature, while equation $III$ ensures the
stability of such a solution. We should note that one can obtain
such conditions by a classical perturbation method \cite{Stab2}.

Here we use these conditions to our de Sitter solutions. In this
model one has
\begin{eqnarray*}
F(R) &=&R-2\Lambda +c_{1}R^{2}, \\
F^{\prime }(R) &=&1+2c_{1}R, \\
F^{\prime \prime }(R) &=&2c_{1}.
\end{eqnarray*}
Substitute de Sitter solution in which $R_{0}=4\Lambda $ ($\Lambda
>0$), we obtain
\begin{eqnarray*}
F(R_{0}) &=&2\Lambda +16c_{1}\Lambda ^{2}, \\
F^{\prime }(R_{0}) &=&1+8c_{1}\Lambda , \\
F^{\prime \prime }(R_{0}) &=&2c_{1}.
\end{eqnarray*}
which confirm that the first and second conditions of Eq.
(\ref{StabCond}) satisfied and for third condition we get
\[
1+\frac{1}{8c_{1}\Lambda }>1,
\]
As a result, de Sitter solution of this model is stable for
$c_{1}>0$.

\subsection{$f(R)=c_{1}R^{2}+K\ln (R)-2\Lambda $ model:}

The cosmological evolution of the $f(R)$ models based on logarithmic
correction has been studied and it has been claimed that such theories have
a well-defined Newtonian limit \cite{Logarithmic}. Here, we desire to
consider $R^{2}$ with logarithmic corrections, simultaneously. To do this,
one can consider $T(R^{\prime })=cR^{\prime }+a_{1}\ln (R^{\prime })+a_{2}$
in Eq. (\ref{Generalf(R)}), which leads to $f(R)=c_{1}R^{2}+K\ln
(R)-2\Lambda $, when
\begin{eqnarray}
c &=&-\frac{1}{8\pi G}  \label{c} \\
a_{1} &=&-\frac{K}{4\pi G},  \label{a1} \\
a_{2} &=&\frac{K+4\Lambda }{8\pi G}.  \label{a2}
\end{eqnarray}%
Considering the mentioned $f(R)$, one can show that Eq. (\ref{Sch}) is a
solution of field equation for
\begin{eqnarray}
\beta ^{\prime \prime } &=&\alpha ^{\prime \prime },  \label{beta2} \\
K &=&\frac{4(\Lambda +12A^{2}\alpha ^{\prime \prime }-3A)}{2\ln (12A)-1}.
\label{K}
\end{eqnarray}%
It is notable that for $k=1$, one may found asymptotic dS solution for
specific value of conformal parameter $\alpha ^{\prime \prime }=\frac{3K}{%
16\Lambda ^{2}}\left[ 2\ln (4\Lambda )-1\right] $.

Now, we search about charged solution. Considering the mentioned model with
Eq. (\ref{K}), it is easy to show that Eq. (\ref{ChargeSolution}) with $%
\beta ^{\prime \prime }=5\alpha ^{\prime \prime }/6$ satisfy the FE (\ref%
{FE1}), provided%
\begin{equation}
c_{1}=\frac{-1}{72}\frac{\Lambda +12\alpha ^{\prime \prime }A^{2}+6A\left[
\ln (12A)-1\right] }{A^{2}\left( 2\ln (12A)-1\right) }.  \label{c11}
\end{equation}

\subsection{$f(R)=c_{1}R^{2}+c_{2}R^{n}-2\Lambda $ model with $n>2$:}

Although most of $f(R)$ gravity models are well-behavior in the weak gravity
regime \cite{Weak}, But for the strong gravity regime, some of these models
have a serious drawback such as singularity problem. In Ref. \cite{Strong},
Kobayashi and Maeda has been resolved the singularity problem arising in the
strong gravity by considering a higher curvature correction term
proportional to $R^{n}$. In order to obtain the mentioned $f(R)$ model ($%
R^{2}$ and $R^{n} $ corrections to Einstein-Hilbert action), one may
consider $T(R^{\prime })=\frac{-1}{8\pi G}R^{\prime }+\frac{(n-2)c_{2}}{8\pi
G}R^{\prime n}+\frac{\Lambda }{2\pi G}$ in Eq. (\ref{Generalf(R)}).

To obtain Schwarzschild solution Eq. (\ref{Sch}) in this model, it is
sufficient to consider $\beta ^{\prime \prime }=\alpha ^{\prime \prime }$
with%
\begin{equation}
c_{2}=\frac{-4(\Lambda +12A^{2}\alpha ^{\prime \prime }-3A)}{(n-2)(12A)^{n}},
\label{c2}
\end{equation}%
which confirm that depend on the sign of $\Lambda $, one may achieve
asymptotic adS/dS solutions provided $\alpha ^{\prime \prime
}=-3c_{2}(n-2)(4\Lambda )^{n-2}$.

Considering $\beta ^{\prime \prime }=5\alpha ^{\prime \prime }/6$ with Eq.( %
\ref{c2}), it is straightforward to obtain charged solution Eq. (\ref%
{ChargeSolution}) for
\begin{equation}
c_{1}=\frac{1}{72}\frac{n\Lambda +12n\alpha ^{\prime \prime }A^{2}-6A(n-1)}{%
A^{2}\left( n-2\right) }.  \label{c111}
\end{equation}%
We should note that the former solution is asymptotically dS (adS) for $%
\Lambda >0$ ($\Lambda <0$) if we set the free parameters of this model in
the following manner%
\begin{eqnarray}
c_{1} &=&-\frac{1}{8\Lambda }+\frac{n\alpha ^{\prime \prime }}{6(n-2)},
\label{c1111} \\
c_{2} &=&-\frac{\alpha ^{\prime \prime }}{3(n-2)(4\Lambda )^{n-2}}.
\label{c22}
\end{eqnarray}

\section{Properties of the exact solutions}

Now, let us look for the singularity of the solutions. Considering
Schwarzschild-(a)dS like (\ref{Sch}) with metric (\ref{Metric}), one can
show that, the Kretschmann scalar is given by%
\begin{equation}
R_{\mu \nu \kappa \eta }R^{\mu \nu \kappa \eta }=24A^{2}+\frac{48M^{2}}{r^{6}%
},  \label{RR1}
\end{equation}
therefore there is a singularity located at $r=0$ (with nonzero $\alpha
=\beta $), as the Schwarzschild solution in general relativity.

\begin{figure}[tbp]
\epsfxsize=8cm \centerline{\epsffile{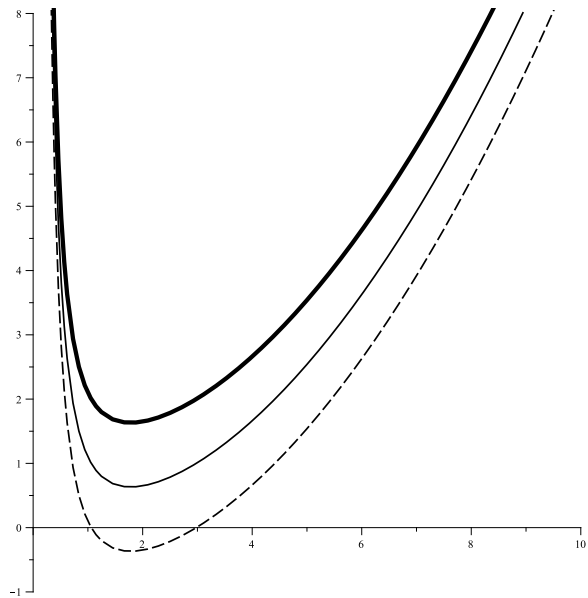}}
\caption{$\protect\lambda(r)$ (Eq. (\protect\ref{ChargeSolution})) versus $r$
for $A=-0.1$, $q=1$, and $k=1$ (bold line), $k=0$ (solid line) and $k=-1$
(dashed line).}
\label{plot}
\end{figure}

In addition, it is easy to show that in the case of charged solution (\ref%
{ChargeSolution}), the Kretschmann scalar is
\begin{equation}
R_{\mu \nu \kappa \eta }R^{\mu \nu \kappa \eta }=24A^{2}+\frac{56q^{4}}{r^{8}%
},  \label{RR2}
\end{equation}
which confirm that the geometry of singularity ($r\longrightarrow 0$) is the
same as Reissner-Nordstr\"{o}m black hole. It is notable that presented
charged black hole solution (\ref{ChargeSolution}), has two horizons for
hyperbolic horizon ($k=-1$) and for $k=0,1$, we encounter with naked
singularity (see Fig. \ref{plot} for more details).

It is notable that the presented solutions are nonsingular for vanishing
mass ($M$) and charge ($q$).

\section{Viability conditions of $F(R)$ theories}

The first study on cosmological viable f(R) models have been
presented in Ref. \cite{segei2006}. The fundamental conditions and
restrictions \cite{Pogosian} that are usually imposed to action
\begin{equation}
S_{g}=\frac{1}{16\pi G}\int d^{4}x\sqrt{-g} F(R)  \nonumber
\end{equation}
theories to provide consistent both gravitational and cosmological
evolutions are:

\begin{enumerate}
\item $F^{\prime \prime }(R)\geq 0$ for $R\gg F^{\prime \prime }(R)$. This
is the stability requirement for a high curvature classical regime \cite%
{Faraoni} and that of the existence of a matter dominated era in
cosmological evolution. A simple physical interpretation can be given to
this condition: if an effective gravitational constant $G_{eff}\equiv
G/(1+F^{\prime }(R))$ is defined, then the sign of its variation with
respect to $R$, $\text{d}G_{eff}/\text{d}R$, is uniquely determined by the
sign of $F^{\prime \prime }(R)$, so in case $F^{\prime \prime }(R)<0$, $%
G_{eff}$ would grow as $R$ does, because $R$ generates more and more
curvature itself. This mechanism would destabilize the theory, as it
wouldn't have a fundamental state because any small curvature would grow to
infinite. Instead, if $F^{\prime \prime }(R)\geq 0$, a counter reaction
mechanism operates to compensate this $R$ growth and stabilize the system.

\item $1+F^{\prime }(R)>0$. This conditions ensures that the effective
gravitational constant is positive, as it can be checked from the previous
definition of $G_{eff}$. It can also be seen from a quantum point of view as
the condition that avoids the graviton from becoming a ghost \cite{Nunez}.

\item $F^{\prime }(R)<0$. Keeping in mind the strong restrictions of Big
Bang nucleosynthesis and cosmic microwave background, this condition ensures
GR behavior to be recovered at early times, that is, $F(R)/R\rightarrow0$
and $F^{\prime }(R)\rightarrow0$ as $R\rightarrow\infty$. Conditions 1 and 2
together demand $F(R)$ to be a monotone increasing function between the
values $-1<F^{\prime }(R)<0$.

\item $F^{\prime }(R)$ must be small in recent epochs. This condition is
mandatory in order to satisfy imposed restrictions by local (solar and
galactic) gravity tests. As the analysis done in \cite{Sawicki} indicates,
the value of $|F^{\prime }(R)|$ must not be bigger than $10^{-6}$ (although
there is still some controversy about this). This is not a needed
requirement if the only goal is to obtain a model that explains cosmic
acceleration.
\end{enumerate}

Now we examine the viability conditions (1-4) for a general $F(R)= R+f(R)$
model described in (7). Explicitly, we have
\begin{eqnarray}
F^{\prime }(R)=2c_1 R+16\pi G R\int^{R}\frac{T(R^{\prime })}{R^{\prime 3}}
dR^{\prime }+8\pi G\frac{T(R)}{R} \\
F^{\prime \prime }(R)=2c_1+8\pi G\frac{T(R)}{R^2}+16\pi G\int^{R}\frac{%
T(R^{\prime })}{R^{\prime 3}}dR^{\prime }
\end{eqnarray}
Validity of the conditions (1-4) depends on the form of the $T(R)$. We will
summarize the different models of F(R)=R+f(R) discussed in this paper as the
follows. Note that since we considered the $F(R)=R+f(R)$ models, thus in
each case we firstly written the explicit for of the F(R).

\subsection{T(R)=T=c.t.e}

In this case, one can obtain
\begin{eqnarray}
F^{\prime }(R)=2c_1 R \\
F^{\prime \prime }(R)=2c_1
\end{eqnarray}
with the following properties:

\begin{enumerate}
\item $F^{\prime \prime }(R)=2c_1 R\geq 0$ for $R\gg 2c_1$. Thus this model
is stable or a high curvature classical regime. The effective gravitational
constant is $G_{eff}\equiv \frac{G}{1+2c_1 R}$. In the early universe epoch,
when $R>>R_0=2c_1$, this effective G reduces to $G_{eff}\simeq \frac{1}{R}$%
,its variation with respect to $R$, $\frac{\text{d}G_{eff}}{\text{d}R}<0$.
Thus $G_{eff}$ would decreases as $R$ grows in the early times,this
mechanism would stabilize the theory, as it would have a fundamental state .

\item Obviously for $R\gg 2c_1$,$1+F^{\prime }(R)>0$. It means that the $%
G_{eff}$ is positive. Thus this case of theory is free from ghosts.

\item $F(R)/R\rightarrow\infty$ and $F^{\prime }(R)\rightarrow\infty$ as $%
R\rightarrow\infty$. Thus only for $c_1=0$, this model recovers GR behavior
at early times.

\item In early times,$F^{\prime }(R)<<$. Thus our model passes the local
solar system tests. Further, from $|2c_1 R|<10^{-6}$ gives the bound on $%
|c_1 |<\frac{10^{-6}}{R},R\approx O(\Lambda)$.
\end{enumerate}

\subsection{$f(R)=-2\Lambda$}

This case describes the Einstein-Hilbert action with a cosmological constant
with
\begin{eqnarray}
F^{\prime }(R)=1 \\
F^{\prime \prime }(R)=0
\end{eqnarray}

\begin{enumerate}
\item $F^{\prime \prime }(R)=1\geq 0$ for $R\in \mathbb{R}$. Thus this model
is stable .

\item Obviously $1+F^{\prime }(R)>0$. It means that the $G_{eff}=G$ is
positive. Thus this case of theory is free from ghosts. Further, it
coincides with GR at the action level.
\end{enumerate}

\subsection{$f(R)=c_{1}R^{2}-2\Lambda $}

Now, we have $F(R)=R+c_{1}R^{2}-2\Lambda$ and thus we've
\begin{eqnarray}
F^{\prime }(R)=1+2c_1 R \\
F^{\prime \prime }(R)=2 c_1
\end{eqnarray}

\begin{enumerate}
\item $F^{\prime \prime }(R)=2 c_1\geq 0$ for $c_1>0$. This is the stability
requirement for a high curvature classical regime of this model and that of
the existence of a matter dominated era in cosmological evolution. This
model induces an effective gravitational constant $G_{eff}\equiv G/2(1+c_1
R) $ is defined, then the sign of its variation with respect to $R$, $\frac{%
\text{d}G_{eff}}{\text{d}R}=-\frac{Gc_1}{2(1+c_1 R)^2}$, is uniquely
determined by the sign of $c_1$, so in case $c_1<0$, $G_{eff}$ would grow as
$R$ does, because $R$ generates more and more curvature itself. The problem
changed when the sign of the $c_1$ reverses. This mechanism would
destabilize the theory, as it wouldn't have a fundamental state because any
small curvature would grow to infinite. Instead, if $c_1\geq 0$, a counter
reaction mechanism operates to compensate this $R$ growth and stabilize the
system.

\item $1+F^{\prime }(R)=2(1+c_1 R)>0$. This conditions ensures that the
effective gravitational constant is positive and from a quantum point of
view as the condition for avoiding from a ghost graviton.

\item $2c_1 R<0$ for $c_1<0$.This condition ensures GR behavior to be
recovered at early times.

\item $F^{\prime }(R)=1+2c_1 R$ must be small in recent epochs. This
condition is mandatory in order to satisfy imposed restrictions by local
(solar and galactic) gravity tests. The value of $|1+2c_1 R|$ must not be
bigger than $10^{-6}$.
\end{enumerate}

\subsection{$f(R)=c_{1}R^{2}+K\ln (R)-2\Lambda $}

In this case the Lagrangian of the model is $F(R)=R+c_{1}R^{2}+K\ln
(R)-2\Lambda $, we have
\begin{eqnarray}
F^{\prime }(R)=1+2c_1 R+\frac{K}{R} \\
F^{\prime \prime }(R)=2 c_1-\frac{K}{R^2}
\end{eqnarray}

\begin{enumerate}
\item $F^{\prime \prime }(R)=2 c_1-\frac{K}{R^2}\geq 0$ for $R\geq\sqrt{%
\frac{K}{2c_1}}$. This is the stability requirement for a high curvature
classical regime of this model and that of the existence of a matter
dominated era in cosmological evolution. This model induces an effective
gravitational constant $G_{eff}\equiv G/(2+2c_1 R+\frac{K}{R})$ is defined,
then the sign of its variation with respect to $R$, $\frac{\text{d}G_{eff}}{%
\text{d}R}=-\frac{G(2 c_1-\frac{K}{R^2})}{(2+2c_1 R+\frac{K}{R})^2}$, is
uniquely determined by the sign of $c_1,K$, so in case $c_1<0,K>0$, $G_{eff}$
would grow as $R$ does, because $R$ generates more and more curvature
itself. The problem changed when the sign of the $c_1,K$ reverses. This
mechanism would destabilize the theory, as it wouldn't have a fundamental
state because any small curvature would grow to infinite.

\item $1+F^{\prime }(R)=2+2c_1 R+\frac{K}{R}>0$. This conditions ensures
that the effective gravitational constant is positive and from a quantum
point of view as the condition for avoiding from a ghost graviton. The root
of the equation $1+F^{\prime }(R)=2+2c_1 R+\frac{K}{R}=0$ locates at $%
R_{\pm}=\,{\frac {-1\pm\sqrt {1-8\,c_{{1}}K}}{4c_{{1}}}}$. This is the
critical curvature. If $c_1>0$ then the model is free from ghosts for $%
R>R_{+},R<R_{-}$. But if $c_1<0$ then the ghosts are present for this range
of the curvature.

\item $1+2c_1 R+\frac{K}{R}<0$ for $c_1<0$ ,$R>R_{+},R<R_{-}$ . This
condition ensures GR behavior to be recovered at early times.

\item $F^{\prime }(R)=1+2c_1 R+\frac{K}{R}$ must be small in recent epochs
only for small negative values of $c_1$. This condition is mandatory in
order to satisfy imposed restrictions by local (solar and galactic) gravity
tests. The value of $|1+2c_1 R+\frac{K}{R}|$ must not be bigger than $%
10^{-6} $.
\end{enumerate}

\subsection{ Model with $f(R)=c_{1}R^{2}+c_{2}R^{n}-2\Lambda$ with $n>2$}

In this case, we have $F(R)=R+c_{1}R^{2}+c_{2}R^{n}-2\Lambda$. Explicitly,
we have
\begin{eqnarray}
F^{\prime }(R)=1+2c_1 R+n c_2 R^{n-1} \\
F^{\prime \prime }(R)=2c_1+n(n-1)c_2 R^{n-2}
\end{eqnarray}
Validity of the conditions (1-4) depends on the signs of the constants $%
c_1,c_2$ for $n>2$. If we take $c_1>0,c_2>0$, then obviously for all values
of the R,$F^{\prime }(R)\geq0,F^{\prime \prime }(R)\geq0$. The conditions
(1,2) are satisfied automatically. But for $c_1<0$ it is possible to have $%
F^{\prime }(R)<0$ . Thus the model satisfy the local solar system tests for
some values of the $c_1,c_2$ . For example take n=3. Then the condition $%
F^{\prime }(R)<0$ can be satisfied by such values of the curvature in the
interval $R_{-}<R<R_{+}$ for $c_2>0$,$c_{1}^2>3c_2$ and $R>R_{+}$ or $%
R<R_{-} $ for $c_2<0$ where $R_{\pm}=-c_1\pm\sqrt{c_{1}^2-3c_2}$.

\subsection{ On the validity of CA coefficients}
 Anomaly is a classical symmetry breaking by quantum
corrections. It seems that by such a definition, this effect has
significance role only in early universe. But now, we show that in a
universe which is in accelerating expansion phase, filled by dark
energy, the phenomena can be described by CA terms. The main idea
belongs to \cite{JHEP2009}. We summarized here the main results (for
details refer to the \cite{JHEP2009} and references there). As we
know that, the CA term, if it is considered as an IR contribution to
the stress-energy, helps to understanding of behavior of the EM
tensor of the matter fields  at low energy of any effective theory.
As we know that, the trace term $<T^{\mu}_{\mu}>$ is of order  of
the $<T^{\mu}_{\mu}>\sim H^4$ where H is the usual Hubble constant.
Cut-off from QCD tells us that this value must be of order
$<T^{\mu}_{\mu}>\sim10^{-12} (eV)^4$ \cite{CC}. This is the
observational vacuum energy density. Difference between these values
is the first cosmological constant problem and has no satisfactory
solution till now. By a systematic analysis of the CA terms
corrections to the vacuum energy density it has been shown that
under certain conditions the CA terms can provide an appropriate
scale observed in nature as predicted by QCD estimates. The first
step is writing the total action for the low energy theory which
contains the usual Einstein-Hilbert action, the conformally
invariant Weyl portion, the higher order curvature invariants and
finally the  anomalous contribution. By writing the total EM tensor
we observe that the resulting expression is correctly proportional
to the equations of motion for some auxiliary fields. It is easy to
show that the anomalous contribution to the stress tensor and the
Weyl one would be of the same order of magnitude. If the spacetime
been conformally flat,then  the Weyl invariant term vanishes, we
have only  the anomalous contribution as the only unique
contribution to the energy tensor. Consider the spacetime as the
static de Sitter. By calculating the finite contribution to the
energy and pressure density, it is shown that the result does not
depend on the CA coefficients. Thus we remove the divergences of the
CA stress tensor. Indeed, the contribution of the CA terms reduces
 to the local, second order in the curvature, geometrical
terms, which are of order $H^4$, and therefore irrelevant at
scales lower than the Planck one.

\section{Conclusion}

It has been shown that $F(R)=R+f(R)$ gravity has (a)dS
Schwarzschild solution. In addition, charged solutions of some
pure $F(R)$ models have been investigated in Ref. \cite{F(R)}. In
the present paper, we considered CA as a source of $F(R)$ gravity
and investigated its consequences.

At first, we used temperature conception of black holes to find a near
horizon solution and then we found an approximate solution which one might
interpret it as a Schwarzschild-(a)dS solution.

Next step is devoted to find exact solutions. Calculations showed that
depending on the values of free parameters, one may find Schwarzschild and
charged solutions. Also, we found that the geometry of uncharged (charged)
singularity ($r \rightarrow 0$) is corresponding to the Schwarzschild
(Reissner-Nordstr\"{o}m) black hole.

Finally, we investigated the fundamental conditions to obtain viable $F(R)$
gravity. In other word, we found some restrictions on free parameters to
guarantee stability as well as ghosts free conditions. In addition, we
checked the behavior of the $F(R)$ theory at early times and also
satisfaction of local (solar and galactic) gravity tests.

It is worthwhile to investigate the thermodynamics properties and also
generalize our solutions to higher dimensional spacetime, and these problems
are left for the future.

\section{acknowledgment}

We are also grateful to Mubasher Jamil and Sergei D. Odintsov for
useful comments on an earlier draft.

\end{document}